\documentclass[aps,prx,twocolumn,showpacs,superscriptaddress,longbibliography]{revtex4-2}
\usepackage{amsmath,amssymb,graphicx}
\usepackage{graphicx,epstopdf}
\usepackage{gensymb}
\usepackage[dvipsnames]{xcolor}
\usepackage[T1]{fontenc}
\usepackage[utf8]{inputenc}
\usepackage{lipsum}
\usepackage{adjustbox}
\newlength\mylength
\newcommand{\be}{\begin{equation}}
\newcommand{\ee}{\end{equation}}
\newcommand{\bea}{\begin{eqnarray}}
\newcommand{\eea}{\end{eqnarray}}
\newcommand{\bse}{\begin{subequations}}
\newcommand{\ese}{\end{subequations}}

\usepackage{multibib}
\usepackage{color}
\usepackage[colorlinks,bookmarks=false,citecolor=darkblue,linkcolor=red,urlcolor=blue]{hyperref}

\definecolor{darkred}{rgb}{0.7,0.0,0.0}

\definecolor{darkblue}{rgb}{0,0.02,0.45}

\definecolor{darkgreen}{rgb}{0.02,0.45,0.0}

\definecolor{violet}{rgb}{0.8,0.2,0.6}

\begin{document}
\title{Ground-state properties of the $S=3/2$ anisotropic triangular lattice antiferromagnet Na$_3$Cr(PO$_4$)$_2$}
\author{A. Magar}
\affiliation{School of Physics, Indian Institute of Science Education and Research Thiruvananthapuram-695551, India}
\author{Sebin J. Sebastian}
\affiliation{School of Physics, Indian Institute of Science Education and Research Thiruvananthapuram-695551, India}
\affiliation{Ames National Laboratory, U.S. DOE, Iowa State University, Ames, IA 50011, USA}
\author{Q.-P. Ding }
\affiliation{Ames National Laboratory, U.S. DOE, Iowa State University, Ames, IA 50011, USA}
\author{Y. Skourski}
\affiliation{High Magnetic Field Laboratory (HLD-EMFL), Helmholtz-Zentrum Dresden-Rossendorf, 01328 Dresden, Germany}
\author{A. A. Tsirlin}
\affiliation{Felix Bloch Institute for Solid-State Physics, Leipzig University, 04103 Leipzig, Germany}
\author{Y. Furukawa}
\email{furukawa@ameslab.gov}
\affiliation{Ames National Laboratory, U.S. DOE, Iowa State University, Ames, IA 50011, USA}
\affiliation{Department of Physics and Astronomy, Iowa State University, Ames, IA 50011, USA}
\author{R. Nath}
\email{rameshchandra.nath@gmail.com}
\affiliation{School of Physics, Indian Institute of Science Education and Research Thiruvananthapuram-695551, India}

\begin{abstract}
We report the crystal structure and magnetic properties of a $S=3/2$ anisotropic triangular lattice compound Na$_3$Cr(PO$_4$)$_2$ employing single-crystal and powder x-ray diffraction, magnetization, heat capacity, and $^{31}$P nuclear magnetic resonance (NMR) experiments, supported by the band structure calculations. Magnetic susceptibility exhibits a broad maximum around 3.5~K, indicating the presence of a short-range antiferromagnetic order, typical of a low-dimensional spin system. 
Magnetization and heat capacity manifest an antiferromagnetic long-range ordering at around $T_{\rm N} \simeq 2.6$~K. This was further confirmed by the drastic NMR line broadening and a peak in the nuclear spin-lattice and spin-spin relaxation rates. The isothermal magnetization data exhibit a field-induced spin-flop transition at around $\mu_0H_{\rm SF} \simeq 1.7$~T reminiscent of an anisotropic two-dimensional magnet, before saturating above $\mu_0H_{\rm sat} \simeq 4.5$~T. The saturation field was further upheld by the field-dependent NMR relaxation measurements at low temperatures. The $^{31}$P NMR spectral shape confirms the commensurate antiferromagnetic nature of the ordering below $T_{\rm N}$. 
\textit{Ab initio} calculations reveal a significant deformation of the triangular spin lattice, resulting in triangles with two antiferromagnetic couplings of similar strength and a much weaker coupling along the third side of the triangle.
\end{abstract}

\maketitle
\section{Introduction}
Since Anderson's inceptive proposal of the resonating valence bond spin-liquid state~\cite{Anderson153}, two-dimensional (2D) $S=1/2$ triangular lattice antiferromagnets (TLAFs) have become the prime focus of research in condensed-matter physics. Even though the antiferromagnetic (AFM) nearest-neighbor (NN) interaction makes the triangular lattice frustrated, the ground state of an isotropic Heisenberg TLAF is calculated to be a chiral 120$\degree$ long-range order (LRO) in both classical and quantum limits~\cite{Bernu134409,Collins605,Capriotti3899}, in agreement with the experiments~\cite{Sakhratov014431,Lee224402}. Nonetheless, 2D TLAFs show nontrivial spin dynamics dominated by magnon decay~\cite{Chernyshev2009} and other many-body effects~\cite{Verresen2019}. 
Further intriguing quantum phases can appear upon introducing unequal exchange couplings in TLAFs. For example,  isosceles triangles lead to a stabilization of a quantum spin liquid (QSL) in an intermediate range of $J'/J$~\cite{Yunoki014408,Weng012407} where $J$ and $J'$ stand for the nearest-neighbor couplings along nonequivalent bonds of the triangle. 

External magnetic field tunes the ground state of a TLAF leading to rich phase diagrams. For example, an up-up-down order manifested by the $\frac13$ magnetization plateau can be stabilized by either thermal or quantum fluctuations~\cite{Chubukov69,Sheng2937,Kawamura1985,Seabra2011}. Other phases, including the V-type, umbrella, and fan-like orders also appear in the phase diagrams alongside the collinear up-up-down phase~\cite{Starykh2014,Yamamoto2014}. Yet another tuning opportunity is offered by an exchange anisotropy that leads to supersolid phases stabilized in the applied field in easy-axis triangular antiferromagnets~\cite{Wang2009,Heidarian2010}.


Discovering materials with triangular networks of magnetic ions facilitates an experimental probe of this interesting physics. In this context, glaserite family with the general formula $A_2A'M(X$O$_4$)$_2$ is particularly interesting because of its high flexibility that allows an accommodation of various transition-metal ions $M$ along with a range of monovalent and divalent cations in the $A$ and $A'$ sites. Depending on the presence of the distortions, the magnetic behavior of glaserite-type compounds ranges from predominantly one-dimensional~\cite{Sebastian064413,Amuneke2207,Tsirlin014401} to two-dimensional with the regular triangular lattice and a varying degree of magnetic anisotropy, as in the spin supersolid candidate Na$_2$BaCo(PO$_4$)$_2$~\cite{Xiang2024} and its Ni analog showing a Bose-Einstein condensation of two-magnon bound states~\cite{Sheng2025}.

\begin{figure}
\includegraphics[width=\columnwidth]{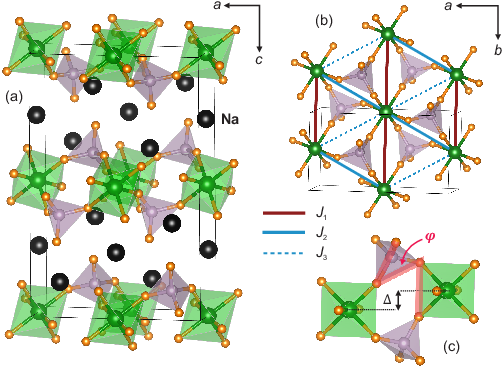}
\caption{\label{Fig1} (a) Crystal structure of Na$_3$Cr(PO$_4)_2$ with the triangular layers stacked along the $c$ direction. (b) Triangular layer with three nonequivalent exchange couplings, $J_1-J_3$. Whereas the Cr--Cr distances of $J_2$ and $J_3$ are identical, the corresponding superexchange pathways are not, as one can see from the positions of the PO$_4$ tetrahedra relative to the Cr--Cr contact. (c) Geometrical parameters of the superexchange pathway: lateral displacement of the octahedra ($\Delta$) and rotation of the PO$_4$ tetrahedron gauged by the dihedral angle $\varphi$.
}
\end{figure}

An interesting subclass of glaserites is formed by purely Na compounds, such as Na$_3$Fe(PO$_4$)$_2$~\cite{Sebastian104425,Ambika015803,Saha094421}. Na$_3$Fe(PO$_4$)$_2$ crystallizes in a monoclinic structure with the $C2/c$ space group and exhibits a slightly anisotropic triangular motif of Fe$^{3+}$ ($S=5/2$) ions. Such a monoclinic distortion splits nearest-neighbor exchange couplings into several groups and gives access to the spatially anisotropic regime of TLAFs. Na$_3$Fe(PO$_4$)$_2$ undergoes a commensurate and collinear AFM ordering at around $T_{\rm N}\sim 10.4$~K and features a spin-flop transition under applied magnetic field before saturating around 35~T~\cite{Sebastian104425}. 
Herein, we report an isostructural compound Na$_3$Cr(PO$_4$)$_2$ with Cr$^{3+}$ featuring a reduced spin of $S=3/2$ compared to its Fe$^{3+}$ counterpart. The Cr compound is also monoclinic (Fig.~\ref{Fig1}) and features a commensurate AFM LRO below $T_{\rm N} \simeq 2.6$~K with a spin-flop transition in an intermediate field, before saturating at around 5~T. A comprehensive phase diagram obtained from thermodynamic and NMR measurements illustrates changes in the magnetic coupling energy. We also uncover a non-trivial deformation of the triangular lattice caused by the varying orientation of the PO$_4$ tetrahedra with respect to the magnetic framework of the glaserite structure.

\section{Methods}
Platelet single crystals of Na$_3$Cr(PO$_4$)$_2$ with a lateral size of 0.1~mm to 0.3~mm were synthesized by a self-flux technique. A mixture of starting materials, Na$_3$PO$_4$, Cr$_2$O$_3$, and NH$_4$H$_2$PO$_4$ in the appropriate molar ratios was mixed and kept in a conical bottom alumina crucible and heated gradually to $900\degree$C. A slow cooling was performed at the rate of $2\degree$C/h till $800\degree$C, and then the sample was allowed to cool naturally to room temperature. The final product yielded a mixture of tiny single crystals and the polycrystalline sample. A few good quality crystals were hand-picked for structural analysis, whereas the remaining part was ground to get the polycrystalline sample used in thermodynamic and NMR measurements. This optimum growth condition was achieved after several attempts by varying the annealing temperature and cooling rate.

Single-crystal x-ray diffraction (XRD) was performed at room temperature using the Bruker KAPPA APEX-II CCD diffractometer equipped with graphite monochromated Mo $K_{\alpha}$ radiation ($\lambda = 0.71073$~\AA). The APEX3 software was used to collect the data, which were further reduced with SAINT/XPREP, followed by an empirical absorption correction using the SADABS program~\cite{Sheldrick1994}. The polycrystalline sample was characterized by powder XRD using a PANalytical Xpert-Pro instrument equipped with Cu~$K_\alpha$ radiation ($\lambda_{av}=1.54182$~\AA) and by the high-resolution powder XRD performed at the ID22 beamline~\cite{Fitch2023} of European Synchrotron Radiation Facility (Grenoble, France) using the wavelength of 0.4\,\r A. 

Magnetization ($M$) measurements were performed as a function of temperature (0.4~K~$\leq T \leq$~380~K) and magnetic field (0~$\leq H \leq$~7~T) using a superconducting quantum interference device (SQUID) (MPMS-3, Quantum Design) magnetometer. Measurements below 1.8~K and down to 0.4~K were carried out using a $^{3}$He attachment to the SQUID magnetometer. High-field magnetization was measured in a pulsed field up to 50~T at the Dresden High Magnetic Field Laboratory~\cite{Tsirlin132407}. Heat capacity ($C_{\rm p}$) as a function of $T$ (2~K~$\leq T \leq$~250~K) and $H$ (0~$\leq H \leq 9$~T) was measured on a small piece of sintered pellet using the relaxation technique in the physical property measurement system (PPMS, Quantum Design).

Nuclear magnetic resonance experiments were conducted using a laboratory-built phase-coherent spin-echo pulse spectrometer within the temperature range 1.5~K~$\leq T \leq$~250~K. These measurements targeted the $^{31}$P nucleus ($I = 1/2$) with the gyromagnetic ratio of $\gamma_{\rm N}/2\pi = 17.236$ MHz/T. While the NMR measurements were carried out at multiple radio frequencies ($\nu$), a detailed temperature-dependent study is presented for $\nu = 17.235$~MHz only, which corresponds to a magnetic field of 1~T. The NMR spectra were obtained by sweeping the magnetic field while maintaining a fixed resonance frequency and utilizing a standard $\pi/2 - \tau - \pi$ pulse sequence. The nuclear spin-lattice relaxation rate ($1/T_1$) was determined by measuring the longitudinal nuclear magnetization as a function of waiting time ($\tau_1$) using the saturation pulse sequence $\pi/2 - \tau_1 - \pi/2 - \tau_2 - \pi$. Similarly, the nuclear spin-spin relaxation rate ($1/T_2$) was extracted by monitoring the decay of spin-echo intensity as a function of the delay time ($\tau_2$) between the $\pi/2$ and $\pi$ pulses.

Density-functional theory (DFT) band-structure calculations were performed in the \texttt{VASP} code~\cite{vasp1,vasp2} using the Perdew-Burke-Ernzerhof flavor of the exchange-correlation potential~\cite{pbe96}. Correlation effects in the Cr $3d$ shell were treated on the mean-field level within DFT+$U$ using the on-site Coulomb repulsion $U_d=5$\,eV, Hund's coupling $J_d=1$\,eV, and double-counting correction in the atomic limit~\cite{Kolay224405}. The results were cross-checked by varying $U_d$ in the range of $3-7$\,eV, which led to a uniform re-scaling of all exchange couplings without any qualitative changes in the magnetic model.

Magnetic parameters were obtained by a mapping procedure~\cite{Xiang2011,Tsirlin2014} for the spin Hamiltonian
\begin{equation}
 \mathcal H=\sum_{\langle ij\rangle}J_{ij}\mathbf S_i\mathbf S_j
\end{equation}
where the summation is over bonds, and $S=3/2$. Magnetization and magnetic susceptibility of this spin Hamiltonian were obtained by quantum Monte-Carlo (QMC) simulations using the \texttt{loop}~\cite{loop} and \verb|dirloop_sse|~\cite{dirloop} algorithms of the \texttt{ALPS} simulation package~\cite{alps} on a 2D finite lattice with up to 144 sites and periodic boundary conditions.

\section{Results and Discussion}
\subsection{Crystal structure}
\begin{figure}
	\includegraphics[width=\columnwidth]{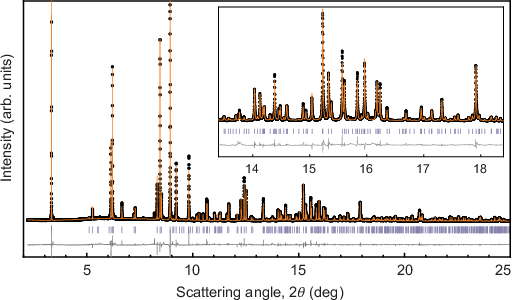}
	\caption{\label{Fig2} Rietveld refinement for the high-resolution XRD data collected at 80\,K. Tick bars show the peak positions, whereas the gray line is the difference curve. The refined lattice parameters are $a=8.99837(2)$\,\r A, $b=4.97809(1)$\,\r A, $c=13.78319(3)$\,\r A, $\beta=90.9559(1)^{\circ}$. The refinement residuals are $R_I=0.036$ and $R_p=0.055$.}
\end{figure}
The crystal structure of Na$_3$Cr(PO$_4$)$_2$ was solved from single-crystal XRD data with direct methods using \texttt{SHELXT-2018/2}~\cite{Sheldrick2015shelxt} and refined by the full matrix least squares on $F^2$ using \texttt{SHELXL-2018/3}, respectively~\cite{Sheldrick2018shelxl}. Details of the room-temperature crystal structure and the refined parameters are summarized in Table~\ref{Table1}. Additionally, we performed Rietveld refinement against the high-resolution powder XRD data collected at 80\,K that confirm the structural model determined from the single crystal (Fig.~\ref{Fig2}). \texttt{JANA2006} was used for the refinement~\cite{jana2006}.

Na$_3$Cr(PO$_4$)$_2$ crystallizes in the monoclinic space group $C2/c$, similar to Na$_3$Fe(PO$_4$)$_2$~\cite{Belkhiria1998,Morozov2001}. The refined atomic positions at room temperature are listed in Table~\ref{Table2}. These refined structural parameters are comparable to those reported for Na$_3$Fe(PO$_4$)$_2$, this is because Cr$^{3+}$ and Fe$^{3+}$ have very similar ionic radii of 0.69\,\r A and 0.63\,\r A, respectively. The Cr$^{3+}$ ions are octahedrally coordinated and are connected into layers via PO$_4$ tetrahedra, with Na$^+$ ions filling the interlayer space. All nearest-neighbor Cr--Cr distances in the triangular magnetic framework are lattice translations of the $C2/c$ space group. Its monoclinic symmetry results in two distinct Cr--Cr distances of 4.991\,\r A ($J_1$) and 5.139\,\r A ($J_2$, $J_3$), see Fig.~\ref{Fig1}b.
\begin{table}[h]
	\caption{Crystallographic data of Na$_3$Cr(PO$_4$)$_2$ at room temperature, obtained from single-crystal XRD.}
	\begin{tabular}{ c   c } 
		\hline\hline
		Empirical formula & Cr Na$_{3}$ O$_8$ P$_2$ \\ 
		Formula weight(M$_r$)  & 1243.64~g~mol$^{-1}$ \\ 
		Temperature & 296(2) K \\ 
		Crystal system  & Monoclinic \\
		Space group  & $C 2/c$ \\
		Lattice parameters &$a=8.984(2)$~\AA~~~$\alpha = 90\degree$\\&$b=4.9908(1)$~\AA~$\beta = 90.830(8)\degree$\\&$c=13.829(4)$~\AA~~~$\gamma = 90\degree$\\	
		Unit cell volume & 620.0(3)~\AA$^3$\\
		$Z$ & 1\\
		Density (calculated) &	3.331~g~cm$^{-3}$\\
		Wavelength & 0.71073 \AA \\
		Radiation type & MoK$\alpha$\\
		Diffractometer & Bruker KAPPA 
		APEX-II CCD\\
		Crystal size & $0.069\times 0.037\times 0.020$~mm$^3$\\
		2$\theta$ range & 2.946 to 24.991$\degree$\\
		Index ranges & $-10\leq h\leq 10$\\& $-5\leq k\leq 5$ \\& $-16\leq l \leq 16$ \\
		$F(000)$ & 604\\
		Reflections collected & 5338\\
		Independent reflections & 542 [$R_{int}$ = 0.0751]\\
		Data / restraints / parameters & 542/24/66\\
		Goodness-of-fit on $F^2$ & 1.249\\
		Final $R$ indices [$I$~$\geq~$2$\sigma$($I$)] &	$R1$ =0.0909,\\& $\omega$$R2$ = 0.2881\\
		$R$ indices(all data) & 	$R1$ = 0.0921,\\& $\omega$$R2$ = 0.2894\\
		Largest diff. peak and hole & 	$1.562 $ and $-1.783$e.\AA$^{-3}$\\
		\hline\hline
	\end{tabular}
	\label{Table1}
\end{table}

\begin{table}
	\caption{Refined atomic parameters of Na$_3$Cr(PO$_4$)$_2$ using single-crystal XRD data at room temperature (upper rows) and high-resolution powder XRD data at 80\,K (lower rows). The Wyckoff positions, atomic coordinates, and isotropic atomic displacement parameters $U_{\rm iso}$(\AA$^2\times 10^2$) are tabulated. In the case of the single crystal, $U_{\rm iso}$ is defined as one-third of the trace of the orthogonalized $U_{ij}$ tensor.}
	\begin{ruledtabular}
	\begin{tabular}{ c c c c c c }
		Atom & Site & $x/a$ & $y/b$ & $z/c$ & $U_{\rm iso}$ \smallskip\\
		Na(1)& $4e$  & $\frac12$ & 0.0471(13) & $\frac34$ & 1.7(2)\\
		     &       & $\frac12$ & 0.0489(3)  & $\frac34$ & 0.78(4) \smallskip\\
		Na(2)& $8f$ &	0.3327(4) & 0.5431(9) &	0.8663(3) & 1.2(1)\\
		     &      & 0.3325(1) & 0.5455(2) & 0.8660(1) & 0.56(3) \smallskip\\ 
		Cr(1)& $4b$ &	$\frac12$ &	0 &	$\frac12$ & 0.6(1)\\
		     &      & $\frac12$ & 0 & $\frac12$ & 0.51(2) \smallskip\\
		P(1)& $8f$ &	0.3302(3) & 0.5228(5) &	0.6110(2) &	0.4(1)\\
		    &      &  0.3300(1) & 0.5221(2) & 0.6108(4) & 0.14(2) \smallskip\\
		O(1)& $8f$ &	0.3901(7)	& 0.3262(1) &	0.5356(5) & 1.0(2)\\
		    &      &  0.3894(2) & 0.3271(3) & 0.5348(1) & 0.09(2) \smallskip\\
		O(2)& $8f$ &	0.4208(7) & 0.7880(1)	& 0.6107(4) &	0.8(2)\\
		    &      &  0.4214(2) & 0.7891(3) & 0.6121(1) & 0.09(2) \smallskip\\
		O(3)& $8f$ &	0.1674(6) & 0.5965(1) &	0.5853(5) & 1.1(2)\\
		    &      &  0.1663(2) & 0.6033(3) & 0.5855(2) & 0.09(2) \smallskip\\
		O(4)& $8f$ &	0.3351(7) & 0.3958(2) & 0.7085(5) &	1.4(2)\\
		    &      &  0.3349(2) & 0.3685(3) & 0.7106(2) & 0.09(2) \\
	\end{tabular}
	\end{ruledtabular}
	\label{Table2}
\end{table}

\subsection{Magnetization and Heat Capacity}
\begin{figure}
\includegraphics[width=\columnwidth]{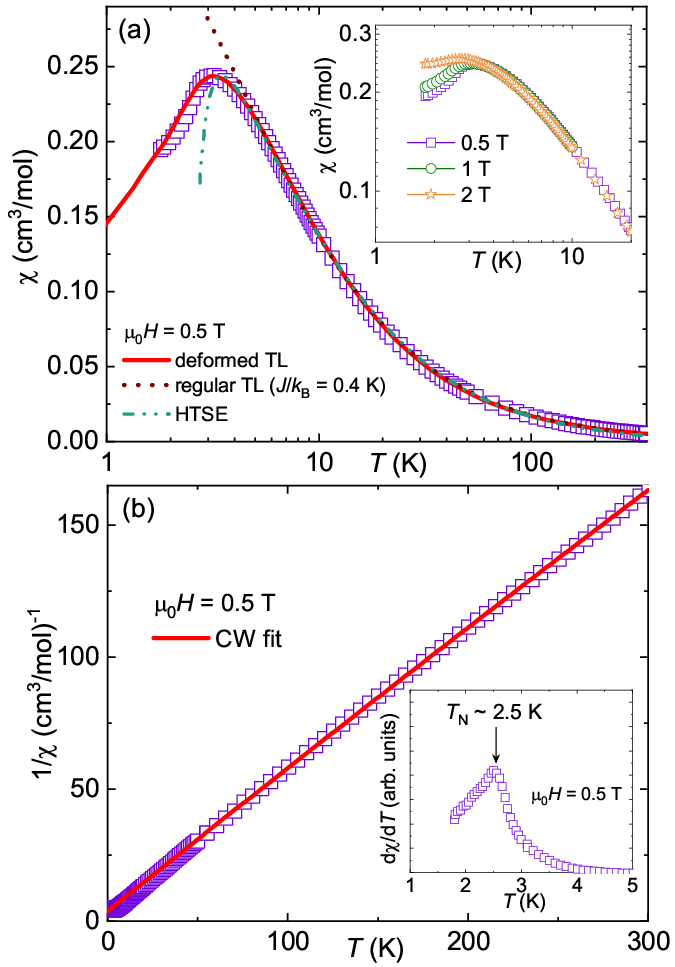}
\caption{\label{Fig3}(a) Magnetic susceptibility ($\chi$) vs $T$ measured in the applied field of 0.5~T. The solid line represents the simulated $\chi(T)$ for the deformed triangular lattice (TL), the dotted line corresponds to the classical Monte-Carlo simulation for a regular TL with $J/k_{\rm B}=0.4$~K, and the dash-dotted line shows the fit using Eq.~\eqref{HTSE}. Inset: $\chi(T)$ measured in different applied fields. (b) $1/\chi$ vs $T$ with the CW fit shown by the solid line. Inset: $d\chi/dT$ vs $T$ measured in 0.5~T pinpointing the transition temperature ($T_{\rm N}$).}
\end{figure}
\begin{figure}
\includegraphics[width=\columnwidth]{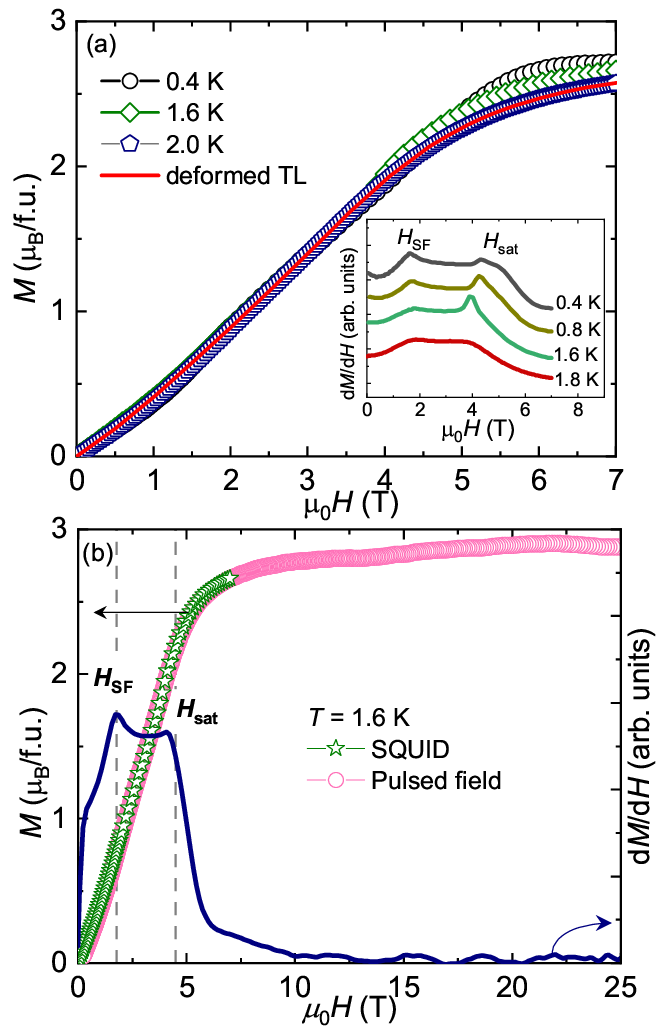}
\caption{\label{Fig4}(a) Magnetic isotherms ($M$ vs $H$) at different temperatures. The solid line represents the calculated magnetization for the deformed TL at $T = 2$~K. Inset: Vertically translated derivative of the magnetization ($dM/dH$) vs $H$ at different temperatures, highlighting the field-induced transition marked by $H_{\rm{SF}}$ and the saturation field $H_{\rm{sat}}$. (b) $M$ and $dM/dH$ vs $H$ measured using pulsed magnetic fields (scaled with respect to the SQUID data) in the left and right $y$-axes, respectively. The sharp features in $dM/dH$ are marked as $H_{\rm{SF}}$ and $H_{\rm{sat}}$.}
\end{figure}
Temperature-dependent magnetic susceptibility $\chi~[\equiv M/H]$ measured in different applied fields is shown in Fig.~\ref{Fig3}(a). $\chi(T)$ increases with a decrease in $T$ and displays a broad maximum around $\sim 3.5$~K. The broad maxima at low temperature are generally observed in low-dimensional magnets due to short-range correlations ~\cite{Sebastian104425,Mohanty184435,Mukharjee224409}. The broad maximum is followed by a slope change at around $T_{\rm N} \simeq 2.5$~K in $\mu_0 H = 0.5$~T, reflecting the transition to a magnetic LRO state. This is highlighted in the $d\chi/dT$ vs $T$ plot shown in the inset of Fig.~\ref{Fig3}(b). With increasing field, $T_{\rm N}$ is suppressed towards low temperatures, typically expected for an AFM ordering.

To analyze the data, $\chi(T)$ can be expressed as a sum of two terms,
\begin{equation}
\label{chi}
    \chi(T)=\chi_0 + \chi_{\rm spin}(T).
\end{equation}
Here, $\chi_0$ is the temperature-independent susceptibility and $\chi_{\rm spin}(T)$ is the intrinsic spin susceptibility. $\chi_{\rm spin}(T)$ can be either the Curie-Weiss (CW) law or the susceptibility obtained from the relevant spin Hamiltonian. First, the data for $T>50$~K were fitted by Eq.~\eqref{chi} using the CW law, $\chi_{\rm spin}(T)=C/(T-\theta_{\rm CW})$. Here, $C$ is the CW constant and $\theta_{\rm CW}$ is the characteristic CW temperature. The fit shown in Fig.~\ref{Fig3}(b) yields $\chi_0 = 1.8(2) \times 10^{-4}$~cm$^3$/mol, $C = 1.8(2)$~cm$^3$K/mol, and $\theta_{\rm CW} = -6.0(6)$~K. From the value of $C$, the effective moment is calculated to be $\mu_{\rm eff} = 3.8(1)~\mu_{\rm B}$ which is close to the theoretically calculated value, $(\mu_{\rm eff})_{\rm theory} = 3.87~\mu_{\rm B}$ for a $S=3/2$ system with $g=2$. The negative value of $\theta_{\rm CW}$ indicates dominant AFM interactions between the Cr$^{3+}$ ions. To estimate the strength of the exchange coupling ($J/k_{\rm B}$) between the Cr$^{3+}$ ions, the mean-field approximation formula, $|\theta_{\rm CW}|=JzS(S+1)/3k_{\rm B}$ is employed. Here, $z$ is the number of nearest neighbors in the triangular plane and $k_{\rm B}$ is the Boltzmann constant. Using $S=3/2$, $z=6$, and $\theta_{\rm CW} = -6.0(6)$~K, we estimated $J/k_{\rm B} = -0.7(1)$~K. Further, the frustration ratio $f=|\theta_{\rm CW}|/T_{\rm N} \simeq 2$ may indicate a weak frustration or low-dimensionality in Na$_3$Cr(PO$_4$)$_2$.

Another possible form of $\chi_{\rm spin}(T)$ is given by the high-temperature series expansion (HTSE) for the $S=3/2$ TLAF model~\cite{Somesh104422}
\begin{eqnarray}
	\begin{split}
		\lefteqn{\chi_{\rm spin}(T) = \frac{N_{\rm A}g^2\mu_{\rm B}^{2}}{k_{\rm B}T} \Big[1.25-9.375x+56.25x^2
		-293.20x^3}\\&+1390.9x^4-6180.33x^5+26172.66x^6
		-106600.27x^7\\&+41952.64x^8
		-1600661.67x^9+5938171.24x^{10}\Big].
	\end{split}
	\label{HTSE}
\end{eqnarray}
Here, $x=J/k_{\rm B}T$, $J$ is the (average) nearest-neighbor exchange coupling, and the expression is valid for $T\geq 8J/k_{\rm B}$. The dash-dotted line in Fig.~\ref{Fig3}(a) represents the fit by Eq.~\eqref{HTSE} above $T > 4$~K. The resulting values of the fitted parameters are, $\chi_0 = 1.8(5)\times 10^{-4}$~cm$^3$/mol, $g=1.96(3)$, and $J/k_{\rm B} = 0.40(3)$~K. However, at $T<6$\,K the TLAF model clearly deviates from the experimental susceptibility, as we show in Sec.~\ref{sec:model} below.

The magnetic isotherms ($M$ vs $H$) measured at different temperatures below $T_{\rm N}$ are shown in Fig.~\ref{Fig4}(a). At $T=0.4$~K, $M$ increases linearly with $H$, shows a slope change around $\mu_0 H_{\rm SF} \sim 1.7$~T indicative of a spin flop (SF) transition, and a tendency of saturation above $\mu_0 H_{\rm sat} = 4.5(3)$~T. In order to clearly visualize these slope changes, the derivative of magnetization with respect to field ($dM/dH$) vs $H$ is plotted in the inset of Fig.~\ref{Fig4}(a) for different temperatures. 
The transition at $H_{\rm SF}$ is pronounced at low temperatures (below $T_{\rm N}$), shifts weakly with temperature, and disappears above 3~K. Figure~\ref{Fig4}(b) shows the pulsed-field data scaled with respect to the SQUID data recorded at $T=1.6$~K. A clear saturation is seen above $\mu_0 H_{\rm sat} \simeq 4.5$~T with the saturation magnetization of $M_{\rm sat} \simeq 2.9$~$\mu_{\rm B}$ close to the expected value $M_{\rm sat}=3 \mu_{\rm B}$ in the fully polarized state of a $S=3/2$ magnet with $g=2$. One can tentatively estimate the saturation field ($H_{\rm sat}$) for a TLAF using the relation $H_{\rm sat} = 9JS/g\mu_{\rm B}$~\cite{Lal014429}. For Na$_3$Cr(PO$_4$)$_2$, $\mu_0 H_{\rm sat} = 4.5(3)$~T corresponds to $J/k_{\rm B} = 0.66(3)$~K, in a reasonable agreement with the results of the $\chi(T)$ analysis.

\begin{figure}
\includegraphics[width=\columnwidth]{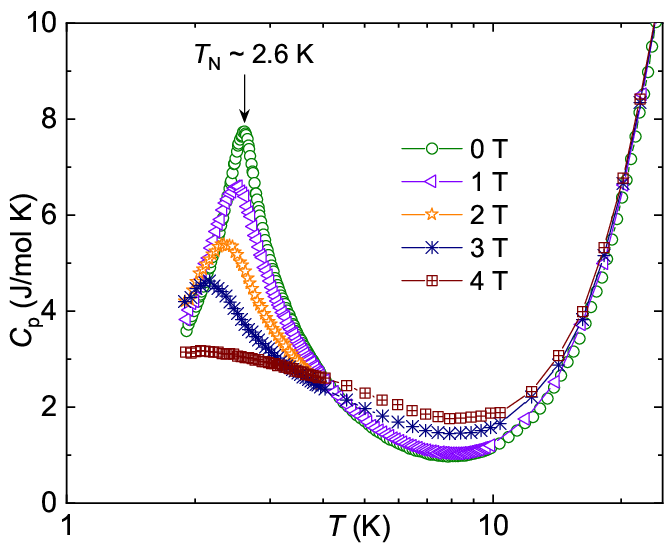}	
\caption{\label{Fig5} (a) Heat capacity ($C_{\rm p}$) vs $T$ in the low-$T$ regime measured in different applied magnetic fields to track down $T_{\rm N}$.}
\end{figure}
Temperature-dependent heat capacity ($C_{\rm p}$) measured in different applied fields is shown in Fig.~\ref{Fig5}. The zero-field $C_{\rm p}(T)$ shows a well-defined sharp anomaly around $T_{\rm N} \simeq 2.6$~K, confirming the magnetic transition. With increasing field, the height of the peak is reduced, and the peak position shifts towards low temperatures, as typically observed for an AFM LRO. For $\mu_0 H > 4$~T, the anomaly is completely suppressed as expected for a fully polarized state. Therefore, the overall $C_{\rm p}$ data are in good agreement with the results of the magnetization measurements.

\subsection{$^{31}$P NMR}
Na$_3$Cr(PO$_4$)$_2$ contains NMR active $^{31}$P nuclei with the nuclear spin $I=1/2$. Further, the PO$_4$ tetrahedra are connected to three adjacent CrO$_6$ octahedra in a triangular motif, as shown in Fig.~\ref{Fig1}, and the crystal structure of Na$_3$Cr(PO$_4$)$_2$ has a unique phosphorus site. Therefore, $^{31}$P is an ideal nucleus for probing magnetic ordering as well as field-induced phases in Na$_3$Cr(PO$_4$)$_2$ via NMR. In the following, the results of temperature and field-dependent NMR measurements are discussed in detail.

\subsubsection{$^{31}$P NMR spectra ($T>T_{\rm N}$)}
\begin{figure}
\includegraphics[width=\columnwidth]{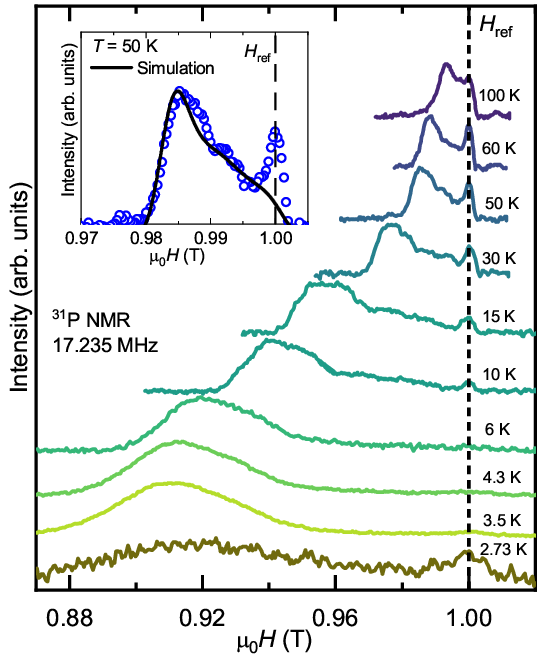}	
\caption{\label{Fig6} Temperature dependence of the $^{31}$P NMR spectra measured at 17.235~MHz, above $T_{\rm N}$. The vertical dashed line marks the $^{31}$P reference resonance field. Inset: Measured spectrum at $T = 50$~K along with the simulated curve (solid line).}
\end{figure}
\begin{figure}
\includegraphics[width=\columnwidth]{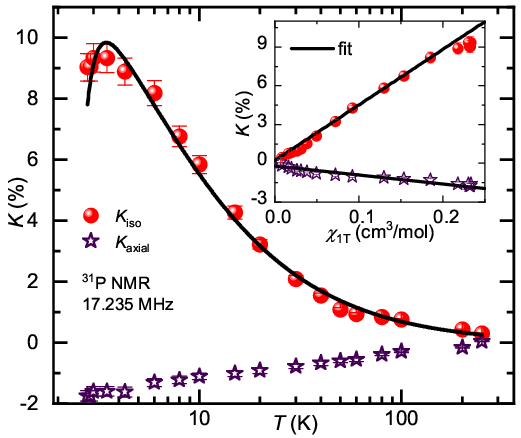}	
\caption{\label{Fig7} Temperature dependence of the NMR shift ($K_{\rm iso}$ and $K_{\rm axial}$) measured at 17.235~MHz and the solid line is the fit to $K_{\rm iso}(T)$ using Eq.~\eqref{NMR_K}. Inset: $K_{\rm iso}$ and $K_{\rm axial}$ vs $\chi$ plot at 1~T, where the solid lines are linear fits.}
\end{figure}
Field-swept NMR spectra measured at $\nu = 17.235$~MHz ($\mu_0 H \simeq 1$~T) above $T_{\rm N}$ are shown in Fig.~\ref{Fig6}. For a better visualization, the spectra at individual temperatures are normalized with the peak amplitude and then vertically translated. The spectra exhibit a single line, as expected for an $I=1/2$ nucleus. With lowering temperature, the line broadens, and the line broadening becomes drastic as one approaches $T_{\rm N}$. A small temperature-independent peak is also visible exactly at the zero-shift position ($H_{\rm ref}$), possibly due to a non-magnetic phosphorous-containing impurity. Further, the peak position of the intrinsic spectra shifts with temperature relative to $H_{\rm ref}$. Initially, the spectra shift to lower fields down to 4~K and then begin shifting to higher fields. The NMR shift ($K$) 
for each temperature is determined by simulating the anisotropic powder spectra using the isotropic ($K_{\rm iso}$) and axial ($K_{\rm axial}$) components of $K$, using the relation~\cite{Sebastian104428,Slichter1992} 
\begin{equation}
K = K_{\text{iso}} + K_{\text{axial}}(3\cos^2\theta - 1),
\label{eq:NMR}
\end{equation}
where $\theta$ defines the orientation of the external field relative to the hyperfine principal axis. One illustrative simulated curve for the spectrum at $T = 50$~K (inset of Fig.~\ref{Fig6}) gives $K_{\rm iso} \simeq 1.08$\% and $K_{\rm axial} \simeq -0.6$\%. Both the components of $K$ ($K_{\rm iso}$ and $K_{\rm axial}$) as a function of $T$ are presented in Fig.~\ref{Fig7}. $K_{\rm iso}(T)$ shows a broad maximum around 3~K, similar to the $\chi(T)$ data.

$K(T)$ is a direct measure of the intrinsic spin susceptibility ($\chi_{\rm spin}$) and is free from any extrinsic contributions. It is expressed in terms of the spin susceptibility $\chi_{\rm spin}$ as
\begin{equation}
    K(T) = K_0 + \frac{A_{\rm hf}}{N_{\rm A}} \chi_{\rm spin}(T),
    \label{NMR_K}
\end{equation}
where $K_0$ is the temperature-independent orbital contribution and $A_{\rm hf}$ is the hyperfine coupling between the $^{31}$P-nucleus and Cr$^{3+}$ spins. We plotted $K$ vs $\chi$ (measured at 1~T) with temperature as an implicit parameter, as shown in the inset of Fig.~\ref{Fig7}. As expected, $K$ and $\chi$ are linearly related, and a fit to the isotropic component using Eq.~\eqref{NMR_K} yields $K_0^{\rm iso} \simeq  0.214$\% and $A_{\rm hf}^{\rm iso} \simeq  0.24$~T/$\mu_{\rm B}$. Similarly, the fit to the axial component gives $K_0^{\rm axial} \simeq -0.20$\% and $A_{\rm hf}^{\rm axial} \simeq -0.04$~T/$\mu_{\rm B}$, indicating a small anisotropic hyperfine field at the $^{31}$P site. The value of $A_{\rm hf}^{\rm iso}$ is comparable to that in other phosphates~\cite{Nagpal17480,Mukharjee224409,Kolay224405}. Since $K$ follows $\chi$, we can thus estimate the exchange coupling between the Cr$^{3+}$ ions by fitting $K_{\rm iso}(T)$ with Eq.~\eqref{NMR_K}, taking $\chi_{\rm spin}(T)$ from Eq.~\eqref{HTSE}. The fit obtained for $T>4$~K by fixing $A_{\rm hf}^{\rm iso} \simeq 0.24$~T/$\mu_{\rm B}$ is shown (solid line) in Fig.~\ref{Fig7}. The fitted parameters are $g = 1.9(6)$ and $J/k_{\rm B}=0.47(3)$~K. This value of $J/k_{\rm B}$ is in good agreement with the one obtained from the $\chi(T)$ analysis.

\subsubsection{$^{31}$P NMR spectra ($T<T_{\rm N}$)}
\begin{figure}
\includegraphics[width=\columnwidth]{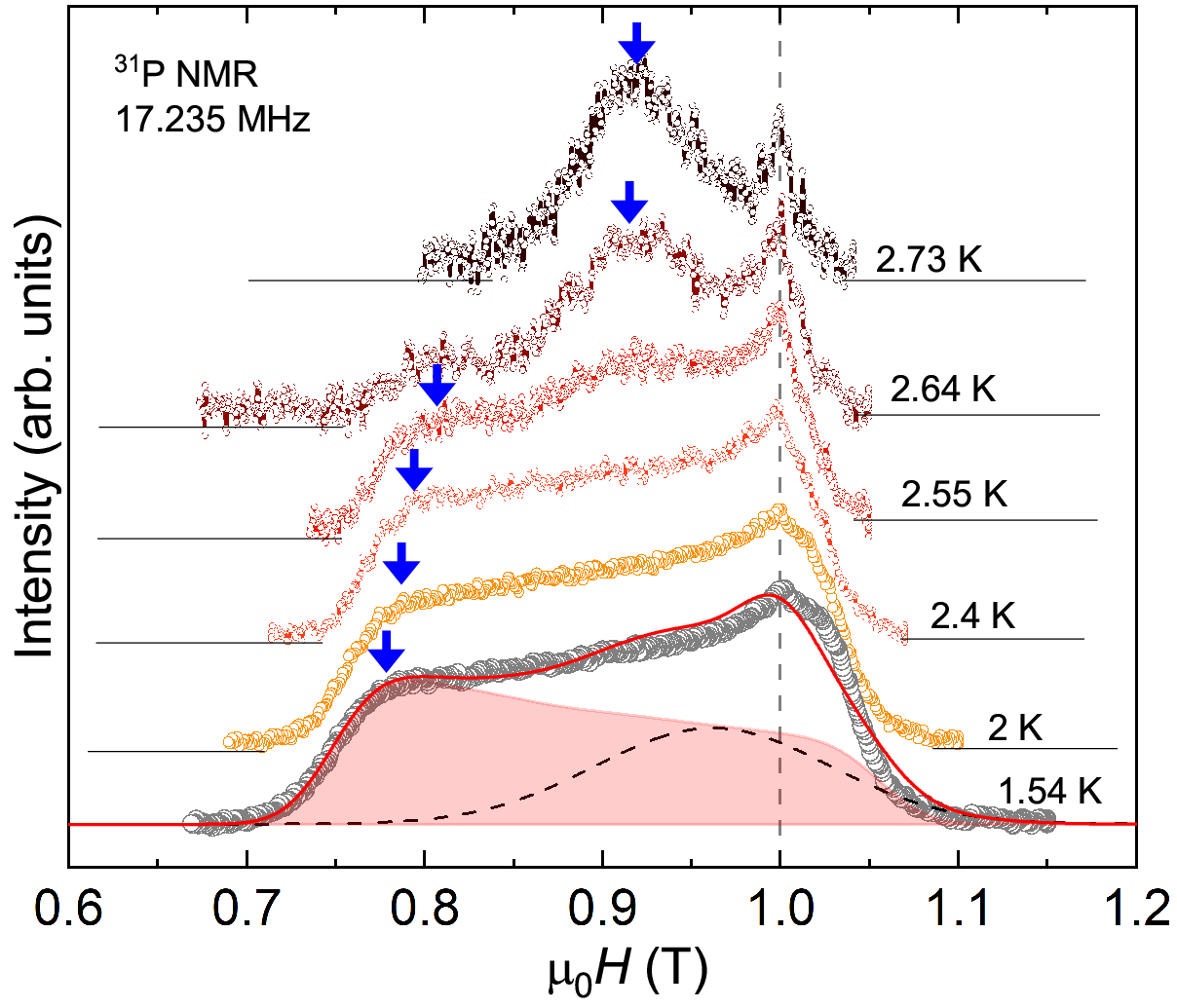}	
\caption{\label{Fig8} $^{31}$P NMR spectra measured below $T_{\rm N}$. The vertical dashed line is the reference (zero-shift) line. The weak peak at the zero-shift position corresponds to a small fraction of the P-containing extrinsic contribution. The area in light-red represents the rectangular-shape of the spectrum at $T = 1.54$~K calculated with  $H_{\rm int}$ = 0.17 T. The dashed line shows the Gaussian-like components of the spectrum. The red solid curve represents the total spectrum, including the aforementioned two components and the impurity contribution. The downward arrows represent the positions where $T_1$ and $T_2$ were measured. 
}
\end{figure}

The $^{31}$P NMR spectra measured below $T_{\rm N}$ are presented in Fig.~\ref{Fig8}. At $T = 2.73$~K (close to $T_{\rm N}$), the line is nearly symmetric but with a sharp peak at the zero-shift position. As one decreases temperature below $T_{\rm N}$, a clear spectral broadening is observed due to the development of the static internal field ($H_{\rm int}$) in the ordered state. Well below $T_{\rm N}$, the spectrum shows a rectangular-like shape with a peak at the zero-shift position.

Typically, for a commensurate AFM ordering, one would expect a rectangular experimental line shape~\cite{Ranjith024422,Ambika015803,Islam174432}, which is due to static internal field in the powder sample~\cite{Yamada1751,Ranjith014415,Kikuchi2660}. 
However, as shown in Fig.~\ref{Fig8}, the shape of the entire experimental spectrum at $T = 1.54$~K cannot be simply reproduced by only a rectangular shape. In addition to the peak around the zero-shift position due to the extrinsic phase, another Gaussian-like component seems to be present, as shown by the black dashed line in Fig.~\ref{Fig8}. While the rectangular line shape is a clear signature of the commensurate AFM ordering, the remaining Gaussian-like line suggests a paramagnetic-like phase fraction existing below $T_N$. Such a paramagnetic-like line could be either due to partial disorder (i.e., staggered component) or spin canting, leading to this inhomogeneous distribution of internal fields~\cite{Bert087203,Nath024431}.


\subsubsection{$^{31}$P spin-lattice relaxation rate ($1/T_1$)}
\begin{figure}
\includegraphics[width=\columnwidth]{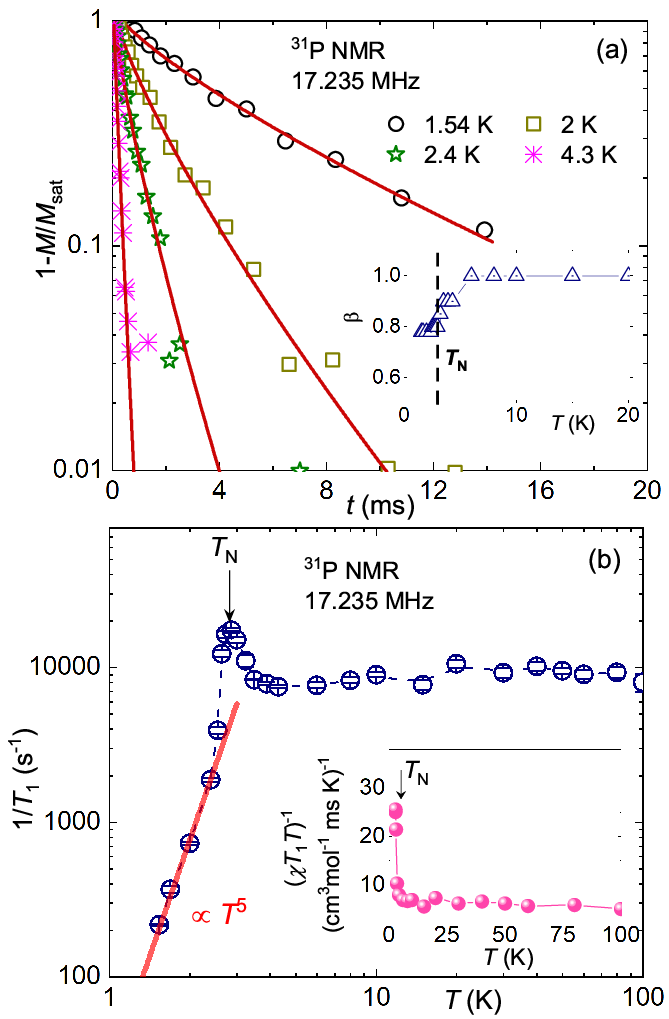}	
\caption{\label{Fig9} (a) Longitudinal nuclear magnetization recovery curves at a few selected temperatures. The solid lines are the fits using Eq.~\eqref{recovery_1/T1}. Inset: temperature dependence of the stretched exponent $\beta$ and the vertical dashed line marks $T_{\rm N}$. (b) Temperature evolution of the $^{31}$P NMR spin-lattice relaxation rate ($1/T_1$). The solid line below $T_{\rm N}$ represents the $T^5$ fit. Inset: temperature dependence of $1/(\chi T_1T)$.}
\end{figure}

The $^{31}$P spin-lattice relaxation rate ($1/T_1$) was  measured at the peak position of the spectra in the paramagnetic state. On the other hand, in the AFM state, 1/$T_1$ was measured at the low-field-edge position of the spectra marked by the arrows in Fig.~\ref{Fig8}. The longitudinal nuclear magnetization recovery curves were fitted by a stretched exponential form~\cite{Mohanty104424},
\begin{equation}
1-\frac{M(t)}{M_{\rm sat}} = A\, e^{-(t/T_1)^\beta}.
\label{recovery_1/T1}
\end{equation}
Here, $M(t)$ is the nuclear magnetization at a time $t$ after the saturation pulse, $M_{\rm sat}$ is the equilibrium nuclear magnetization, and $\beta$ is the stretched exponent that accounts for a distribution of relaxation times. In Fig.~\ref{Fig9}(a), a few recovery curves [$(1-\frac{M}{M_{\rm sat}})$ vs $t$] along with the fits using Eq.~\eqref{recovery_1/T1} are shown. $\beta$ as a function of $T$ is plotted in the inset of Fig.~\ref{Fig9}(a). At high temperatures, $\beta \simeq 1$ (i.e., single-exponential) and is nearly temperature independent as expected for an $I=1/2$ nucleus. This also suggests a homogeneous relaxation in the paramagnetic region ($T>T_{\rm N}$). The value of $\beta$ drops below unity as one approaches $T_{\rm N}$, indicating the distribution of relaxation times~\cite{Mohanty184435}.

The obtained $1/T_1$ is presented in Fig.~\ref{Fig9}(b). At high temperatures, it is almost temperature independent, as expected due to paramagnetic fluctuations. With lowering temperature, $1/T_1$ starts increasing below 4~K, and exhibits a sharp peak at around $T_{\rm N} \simeq 2.6$~K, confirming a critical slowing down of spin fluctuations and a transition to a LRO state~\cite{Sebastian064413}. This peak value is close to the observed $T_{\rm N} \simeq 2.5$~K in the $C_{\rm p}(T)$ data under a magnetic field of 1 T. 
Below $T_{\rm N}$, $1/T_1$ drops sharply, a clear indication of the relaxation due to scattering of magnons by the nuclear spins. For $T\ll \Delta/k_{\rm B}$, the temperature dependence of $1/T_1$ in the ordered state should follow either a $T^3$ or $T^5$ behaviour due to a two-magnon or a three-magnon Raman process, respectively. Here, $\Delta/k_{\rm B}$ is the energy gap in the spin-wave spectrum. As shown in Fig.~\ref{Fig9}(b) (solid line), $1/T_1$ below $T_{\rm N}$ follows a $T^5$ behavior, indicative of a three-magnon Raman process~\cite{Ranjith014415,Yogi024413}. 
It is to be noted that we also measured $T_1$ at the central position of the spectra in the AFM state and found no obvious position difference in $T_1$.

To elucidate the role of spin fluctuations above $T_{\rm N}$, we examined the temperature evolution of $1/(\chi T_1 T)$, shown in the lower inset of Fig.~\ref{Fig9}(b). The nuclear spin-lattice relaxation rate normalized by temperature, $1/T_1T$, provides a direct measure of low-energy spin dynamics and is connected to the imaginary part of the dynamical susceptibility, $\chi^{\prime\prime}_M(\vec{q},\omega_{\rm N})$, as~\cite{Moriya23,Moriya516}
\begin{equation}
\frac{1}{T_{1}T} = \frac{2\gamma_{\rm N}^{2}k_{\rm B}}{N_{\rm A}^{2}}
\sum\limits_{\vec{q}}\mid A(\vec{q})\mid
^{2}\frac{\chi^{\prime\prime}_{\rm M}(\vec{q},\omega_{\rm N})}{\omega_{\rm N}}.
\label{t1form}
\end{equation}
Here, the summation runs over all wave vectors $\vec{q}$ within the first Brillouin zone, $\omega_{\rm N}$ is the measured NMR frequency, and $A(\vec{q})$ represents the hyperfine form factor. In the limit $\vec{q}=0$ and $\omega_{\rm N}\rightarrow 0$, the real part of the dynamical susceptibility $\chi_{\rm M}(\vec{q},\omega_{\rm N})$ reduces to the uniform static susceptibility $\chi$. Consequently, a nearly temperature-independent behavior of $1/(\chi T_1 T)$ is expected at high temperatures ($T > \theta_{\rm CW}$). Indeed, as shown in the lower inset of Fig.~\ref{Fig9}(b), $1/(\chi T_1 T)$ is temperature-independent just above $T_{\rm N}$. Upon cooling, a pronounced increase in $1/(\chi T_1 T)$ occurred below about $\sim 5$~K signaling dominant correlations due to $\vec{q} \neq 0$ as the magnetic LRO sets in~\cite{Nath214430}. Moreover, the enhancement of $1/(\chi T_1 T)$ over a narrow temperature range just above $T_{\rm N}$ reflects only a weak frustration as expected from the small frustration ratio $f\simeq 2$.

\subsubsection{$^{31}$P spin-spin relaxation rate ($1/T_2$)}
\begin{figure}
\includegraphics[width=\columnwidth]{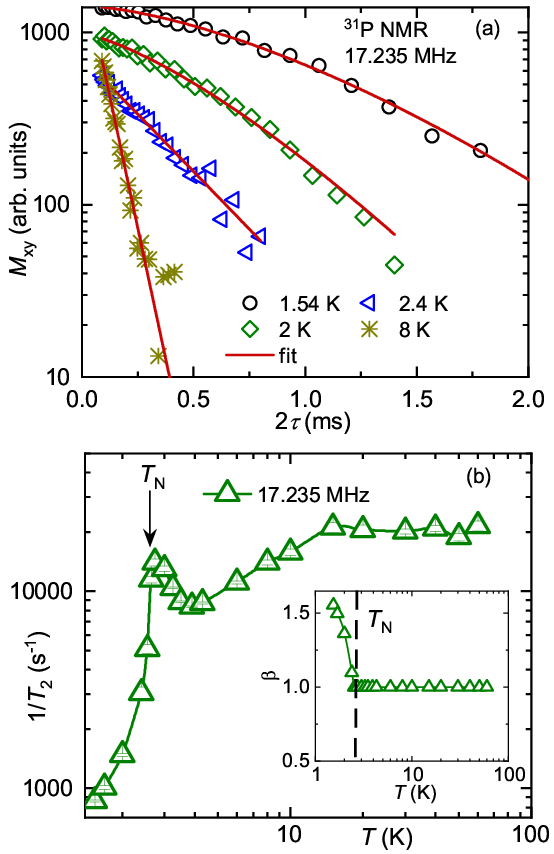}	
\caption{\label{Fig10} a) Transverse magnetization recovery curves at a few selected temperatures, with the solid lines showing fits as described in the text. (b) Temperature evolution of the $^{31}$P NMR spin-spin relaxation rates ($1/T_2$). Inset: Temperature dependence of the exponent $\beta$.}
\end{figure}
The measurement of the nuclear spin-spin relaxation rate $1/T_2$ was done by monitoring the decay of the transverse magnetization ($M_{xy}$) after a $\pi/2-\tau-\pi$ pulse sequence as a function of the pulse separation time $\tau$. The decay curves were then fitted by
\begin{equation}
M_{xy} = M_0 e^{(-{2\tau}/{T_2})^\beta}.
\label{recovery_1/T2}
\end{equation}
Figure~\ref{Fig10}(a) shows the fits for a few selected temperatures. The extracted $1/T_2$ and $\beta$ are plotted as a function of temperature in Fig.~\ref{Fig10}(b). The temperature-dependent $1/T_2$ features nearly the same behavior as $1/T_1$. A sharp peak in $1/T_2$ at around $T_{\rm N}\simeq 2.5$~K under 1~T is followed by a smooth decay below $T_{\rm N}$.
The value of $\beta$ is nearly 1 and temperature-independent at higher temperatures [inset of Fig.~\ref{Fig10}(b)] and then increases rapidly below $T_{\rm N}$.  A similar behavior has been reported in the two-dimensional triangular lattice compound Na$_3$Fe(PO$_4$)$_2$~\cite{Ambika015803}.

\subsubsection{Field dependence of the $^{31}$P NMR spectra, $1/T_1$, and $1/T_2$}
\begin{figure}
\includegraphics[width=\columnwidth]{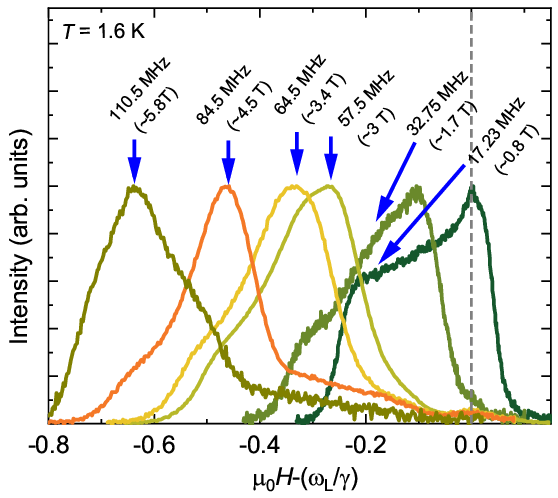}	
\caption{\label{Fig11} $^{31}$P NMR spectra measured at different frequencies at $T=1.6$~K ($T < T_{\rm N}$). The $x$-axis is corrected for the Larmor field in order to highlight the spectral shift. The vertical dashed line represents the zero-shift position. The downward arrows point to the positions where $T_1$ and $T_2$ were measured.}
\end{figure}

To detect the field-induced transition as well as the saturation field, we recorded the NMR spectra, $1/T_1$, and $1/T_2$ at $T=1.6$~K~$(T<T_{\rm N}$) at different frequencies well above and below $H_{\rm sat}$. As evident from Fig.~\ref{Fig11}, for the low frequency of $\sim 17.235$~MHz, the spectrum roughly exhibits rectangular shape as discussed above. With increasing frequency (or magnetic field), the spectral shape evolves continuously and the center of gravity moves away from the zero-shift position. In the intermediate fields, the spectra could be fitted with two components (rectangular and Gaussian-like). The spectral weight associated with the Gaussian-like component increases, whereas the weight associated with the rectangular one decreases upon increasing field. At higher fields, $H > H_{\rm sat}$ (e.g. at 110.5~MHz), the spectrum acquires a characteristic shape, which is similar to that observed in the paramagnetic state, reflecting nearly uniform Cr$^{3+}$ magnetic moments pointing in the external field direction in the fully polarized state. From the shift ($\Delta H$) of the NMR spectrum measured at 110.5~MHz with respect to the zero-shift position, we estimated the saturation moment utilizing the hyperfine coupling constant $A_{\rm hf}^{\rm iso}$ to be $M_{\rm sat} \sim \Delta H/A_{\rm hf}^{\rm iso} \sim 0.62/0.24 \sim 2.6~\mu_{\rm B}$/Cr$^{3+}$. This is in a good agreement with $M_{\rm sat}$ obtained from the $M(H)$ measurements [see Fig.~\ref{Fig4}(a)]. 
\begin{figure}
\includegraphics[width=\columnwidth]{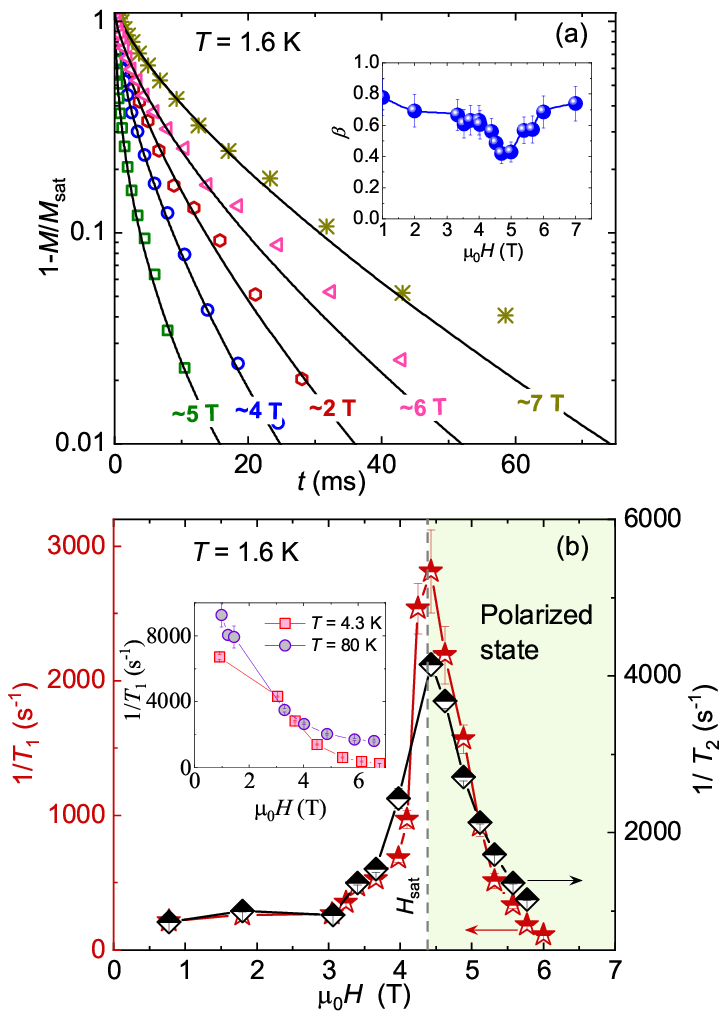}	
\caption{\label{Fig12} (a) Longitudinal nuclear magnetization recovery curves measured at different resonance fields and at a fixed temperature of $T = 1.6$~K along with the fits (solid lines). Inset: magnetic field dependence of $\beta$. (b) Field evolution of $1/T_1$ and $1/T_2$ at $T=1.6$~K in the left and right $y$-axes, respectively. Inset: $1/T_1$ vs $H_{\rm peak}$ in the paramagnetic state measured at $T=4.3$~K and $80$~K.}
\end{figure}

In most cases, the longitudinal nuclear magnetization recovery curves at $T=1.6$~K were measured at the fields corresponding to the peak position, but in low magnetic fields, we measured those at the edge position or near the center of the spectrum as shown by the downward arrows in Fig.~\ref{Fig11}. The corresponding recovery curves are presented in Fig.~\ref{Fig12}(a).  
The recovery curves follow Eq.~\eqref{recovery_1/T1} and the obtained $1/T_1$ as a function of magnetic field is shown in Fig.~\ref{Fig12}(b). The magnetic field dependence of $\beta$ is shown in the inset of  Fig.~\ref{Fig12}(a).

 While $1/T_1$ exhibits a smooth decay with magnetic field in the paramagnetic state ($T > T_{\rm N}$) as shown in the inset of Fig.~\ref{Fig12}(b), a pronounced peak is observed at $\mu_0 H  \simeq 4.5$~T for $T <T_{\rm N}$ ($T=1.6$~K) [Fig.~\ref{Fig12}(b)] that matches with $H_{\rm sat}$. In the polarized state, $1/T_1$ is expected to scale with $(dM/dH)T$ as in the paramagnetic state. Therefore, the peak near $H_{\rm sat}$ in $1/T_1$ correspond to the maximum in the $dM/dH$ vs $H$ curve [inset of Fig.~\ref{Fig4}(a)]. At low magnetic fields below $H_{\rm sat}$, however, the relaxation is dominated by magnon scattering in the AFM ordered state rather than the paramagnetic-like fluctuations, and thus a simple proportionality between $1/T_1$ and $dM/dH$ is not envisaged. Hence, the peak at $H_{\rm sat}$ signals the transition from the AFM ordered state to a fully polarized state, in good agreement with the magnetization data~\cite{Sakhratov014431}. Even $\beta$ as a function of $H$ also features an anomaly at $H_{\rm sat}$. Notably, $1/T_2$ also features a distinct peak at $H_{\rm sat}$, coinciding with the $1/T_1$ data [Fig.~\ref{Fig12}(b)]. Unfortunately, no clear feature is detected at $H_{\rm SF}$. Moreover, the absence of any anomaly in the intermediate fields would also rule out the presence of a $1/3$-plateau in this material~\cite{Koutroulakis024410}.

\subsection{Microscopic magnetic model}
\label{sec:model}

Prior to calculating exchange couplings in Na$_3$Cr(PO$_4)_2$, it is instructive to analyze the symmetry of this material. In contrast to the ideal glaserite structure, Na$_3$Cr(PO$_4)_2$ does not feature three-fold rotational symmetry that would render the bonds of the triangular lattice equivalent. Two distinct Cr--Cr distances of $2\times 4.991$\,\r A ($J_1$) and $4\times 5.139$\,\r A ($J_2$, $J_3$) are observed instead. A more subtle aspect of this distorted structure is that even the bonds with the same Cr--Cr distances of 5.139\,\r A are not symmetry-related. The triangular layers are located in the $z=0$ and $z=\frac12$ planes (Fig.~\ref{Fig1}a), whereas the two-fold symmetry axes of the $C2/c$ space group are at $z=\frac14$ and $z=\frac34$. Therefore, both the two-fold rotational axis and the $c$ glide plane connect the adjacent triangular layers but do not impose any symmetry within the plane. The symmetry of each triangular layer alone is restricted to the diagonal translations due to the $C$-centering. This renders $J_2$ and $J_3$ two distinct couplings.

\begin{table}
\caption{\label{tab:exchange}
Nearest-neighbor exchange couplings in Na$_3$Cr(PO$_4)_2$ calculated with DFT+$U$ ($U_d=5$\,eV) and the relevant geometrical parameters: Cr--Cr distance ($d$), lateral displacement of the CrO$_6$ octahedra ($\Delta$), and rotation angle of the PO$_4$ tetrahedron ($\varphi$), see also Fig.~\ref{Fig1}. Room-temperature structural data are used.
}
\begin{ruledtabular}
\begin{tabular}{c@{\hspace{2em}}ccc}
                 & $J_1$ & $J_2$ & $J_3$ \\
 $J$ (K)         &  1.3  &  1.6  & $-0.4$ \\
 $d$ (\r A)      & 4.991 & 5.139 & 5.139 \\
 $\Delta$ (\r A) &  1.09 &  0.74 &  1.10 \\
 $\varphi$ (deg) & 141.0 & 153.4 & 124.5 \\
\end{tabular}
\end{ruledtabular}
\end{table}

Table~\ref{tab:exchange} lists the $J_1-J_3$ values obtained from DFT. A major difference between $J_2$ and $J_3$ is observed, despite the equal Cr--Cr distances of these two couplings. This difference can be attributed to the geometry of the double PO$_4$ bridges that connect the CrO$_6$ octahedra and mediate superexchange. We quantify it using two parameters, the lateral displacement of the two octahedra ($\Delta$) and the angle $\varphi$ showing the rotation of the PO$_4$ tetrahedron with respect to the oxygen plane of the superexchange pathway (Fig.~\ref{Fig1}c). Both $\Delta$ and $\varphi$ are notably different between $J_2$ and $J_3$, thus illustrating the distinct nature of these couplings, which is also visible from the top view of the triangular layer in Fig.~\ref{Fig1}b.

Previous studies~\cite{Roca1998,Nath014407,Mukharjee2020} suggest that increasing $\Delta$ suppresses AFM exchange. This is indeed the case if one considers $\Delta_2<\Delta_3$, although $\Delta_3$ is comparable to $\Delta_1$, despite $J_1$ being AFM and $J_3$ being ferromagnetic. A better correlation is found for the relative orientation of the PO$_4$ tetrahedron. Increasing the $\varphi$ angle, namely, bringing an apical oxygen of the tetrahedron closer to the plane of the superexchange pathway, facilitates AFM exchange. Such a non-trivial dependence of superexchange on the rotation of the tetrahedral polyanion was previously reported in Cu$^{2+}$ quantum magnets~\cite{Tsirlin2010,Tsirlin014401}.

Turning to the experimental data, we note that the ideal TLAF model of Eq.~\eqref{HTSE} describes the susceptibility down to 6\,K, but extending the fit toward lower temperatures using classical Monte-Carlo simulations leads to an increasing deviation from the experimental curve. A much better description of the susceptibility is obtained by taking into account the deformation of the triangular lattice. Choosing $J_1=0.54$\,K, $J_2=0.67$\,K, and $J_3=-0.4$\,K along with $g=1.91$, we arrive at an excellent fit of the experimental susceptibility down to $T_N$ (Fig.~\ref{Fig3}) and also reproduce the field-dependent magnetization (Fig.~\ref{Fig4}). The fitted $J$ values are generally consistent with the DFT results, although $J_1$ and $J_2$ are somewhat smaller in magnitude, probably due to inaccuracies of DFT in the evaluation of weak exchange couplings.

The 2D model given by $J_1-J_3$ lacks LRO at a finite temperature as per Mermin-Wagner theorem. DFT calculations return a minute interlayer coupling $J_{\perp}\simeq 0.05$\,K, such that $J_{\perp}/J_2\simeq 0.03$. This weak coupling could nevertheless stabilize LRO. Alternatively, a small single-ion anisotropy of the same magnitude results in $T_N=2.4$\,K within the 2D model, in an excellent agreement with the experimental value of 2.5\,K. We thus conclude that Na$_3$Cr(PO$_4)_2$ is a quasi-2D antiferromagnet where individual triangular planes support collinear order, whereas LRO is driven by weak sub-leading terms, such as single-ion anisotropy or interlayer couplings.

\section{Discussion and Summary}
\begin{figure}[h]
\includegraphics[width=\columnwidth]{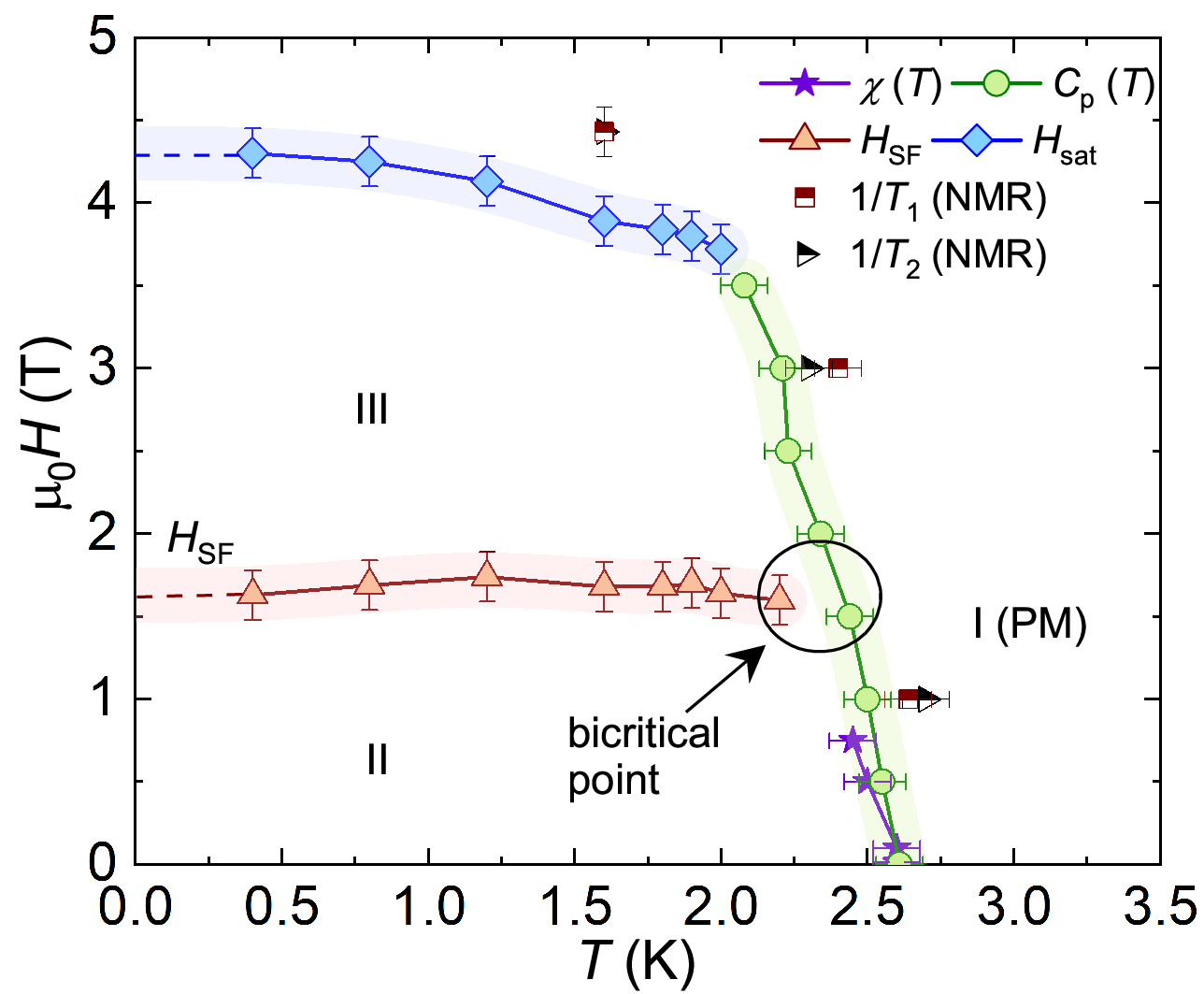}	
\caption{\label{Fig13} The $H - T$ phase diagram of Na$_3$Cr(PO$_4$)$_2$ constructed using the $\chi(T)$, $M(H)$, $C_{\rm p}(T)$, and NMR data. The SF field $H_{\rm SF}$ and bicritical point are marked.}
\end{figure}
To visualize different experimentally realized phases in Na$_3$Cr(PO$_4$)$_2$, a $H-T$ phase diagram is presented in Fig.~\ref{Fig13} by extracting the transition temperature and critical fields from $\chi(T)$, $M(H)$, and $C_{\rm p}(T)$ data, as well as the $^{31}$P NMR relaxation rates. The data points from different probes agree within the experimental error bars. The phase diagram features three distinct phases. The high-temperature paramagnetic phase (I) is separated from the ordered AFM phases by a phase boundary traced by $T_{\rm N}$. Application of magnetic field not only suppresses $T_{\rm N}$ towards low temperatures but also induces a SF transition toward phase III at $H_{\rm SF}$. As the field is increased beyond 4.5~T, the spin system becomes fully polarized.

An ideal TLAF should feature $120^{\circ}$ order in zero field and at least two field-induced phases, including the $\frac13$ plateau. The field-temperature diagram obtained in our work is clearly different. In fact, it is more reminiscent of a two-sublattice collinear antiferromagnet. The observation of the rectangular NMR spectral shape in the LRO state is also indicative of the collinear magnetic structure, although an additional Gaussian-like component was detected in the spectrum as well. The origin of this component requires further dedicated investigation.

From the microscopic standpoint, the collinear order is rationalized by the strong deformation of the triangular spin lattice. Remarkably, this deformation originates not from the different Cr--Cr distances for $J_1$ vs. $J_2$ and $J_3$, but rather from the difference between $J_2$ and $J_3$ themselves. With $J_3$ turning ferromagnetic, frustration is released, and collinear order ensues. The mechanism of the deformation pertains to a rather inconspicuous geometrical difference in the double bridges of the PO$_4$ tetrahedra that are responsible for the magnetic couplings between disconnected Cr$^{3+}$ ions. It illustrates the strong sensitivity of magnetic superexchange to the positions and orientations of polyanions~\cite{Janson2014,Nekrasova2020,Mukharjee224409} and necessitates a careful microscopic analysis. The difference between $J_2$ and $J_3$ may also occur in Na$_3$Fe(PO$_4)_2$ where these two couplings were previously considered equivalent~\cite{Sebastian104425,Ambika015803,Saha094421}. 

To summarize, we probed the static as well as dynamic properties of the  $S=3/2$ anisotropic triangular lattice AFM compound Na$_3$Cr(PO$_4$)$_2$. The crystal structure of Na$_3$Cr(PO$_4$)$_2$ solved from the single-crystal XRD data resembles Na$_3$Fe(PO$_4$)$_2$ with only a slight difference in the bond lengths. Magnetization data reveal the emergence of a short-range AFM order below 3.5~K and an AFM LRO below $T_{\rm N} \simeq 2.6~K$. Isothermal magnetization exhibits a spin-flop transition at around $\mu_0H_{\rm SF} \simeq 1.7$~T and saturation above $\mu_0H_{\rm sat} \simeq 4.5$~T at $T=1.6$~K. A sharp anomaly in the heat capacity confirms the onset of an AFM LRO below $T_{\rm N}$. 
NMR data confirm the formation of LRO and suggest its collinear nature, albeit with an additional Gaussian-like contribution of unknown origin. Our microscopic analysis reveals a large deformation of the triangular lattice that stabilizes collinear magnetic order in this material.

\acknowledgments
We would like to acknowledge SERB, India bearing sanction Grant No.~CRG/2022/000997. We also acknowledge the support of HLD-HZDR, member of the European Magnetic Field Laboratory (EMFL), and the synchrotron beamtime provided by ESRF. SJS acknowledges the Fulbright-Nehru Doctoral Research Fellowship Award No.~2997/FNDR/2024-2025 and the Prime Minister’s Research Fellowship (PMRF) scheme, Government of India, to be a visiting research scholar at the Ames National Laboratory. The research was supported by the US Department of Energy, Office of Basic Energy Sciences, Division of Materials Sciences and Engineering. Ames National Laboratory is operated for the US Department of Energy by Iowa State University under Contract No.~DEAC02-07CH11358. The work in Leipzig was funded by funded by the Deutsche Forschungsgemeinschaft (DFG, German Research Foundation) -- TRR 360 -- 492547816 (subproject B3). AT gratefully acknowledges the computing time made available to him on the high-performance computer at the NHR Center of TU Dresden. This center is jointly supported by the Federal Ministry of Education and Research and the state governments participating in the NHR [www.nhr-verein.de/unsere-partner].


%

\end{document}